\documentclass[aps,prc,nofootinbib,twocolumn,superscriptaddress,showpacs,
floatfix]{revtex4}
\usepackage{mathrsfs}
\usepackage{latexsym}
\usepackage{amsmath}
\usepackage{amssymb}
\usepackage{graphicx}
\usepackage{longtable}

\usepackage{bbm}
\usepackage{epsfig}
\usepackage[usenames]{color}
\usepackage[colorlinks=true,citecolor=blue,linkcolor=blue]{hyperref}

\begin{document}

\title{Quark-to-Gluon composition of the Quark-Gluon Plasma in relativistic heavy-ion collisions}

\author{F. Scardina}
\affiliation{INFN-Laboratori Nazionali del Sud, Via S. Sofia 62, I-95123
Catania, Italy}
\affiliation{Physics Dept., University of Messina, Italy}

\author{M. Colonna}
\affiliation{INFN-Laboratori Nazionali del Sud, Via S. Sofia 62, I-95123
Catania, Italy}

\author{S. Plumari}
\affiliation{INFN-Laboratori Nazionali del Sud, Via S. Sofia 62, I-95123
Catania, Italy}
\affiliation{Physics and Astronomy Dept., University of Catania, Italy}

\author{V. Greco}
\affiliation{INFN-Laboratori Nazionali del Sud, Via S. Sofia 62, I-95123
Catania, Italy}
\affiliation{Physics and Astronomy Dept., University of Catania, Italy}


\begin{abstract}
We study the evolution of the quark-gluon composition of the plasma created in ultra-Relativistic Heavy
Ion Collisions (uRHIC's) employing a partonic transport theory that includes both elastic and inelastic collisions plus a mean fields dynamics associated to the widely used quasi-particle model. 
The latter, able to describe
lattice QCD thermodynamics, implies a "chemical" equilibrium ratio between quarks and gluons
strongly increasing as $T\rightarrow T_c$, the phase transition temperature. Accordingly we see in
realistic simulations of uRHIC's a rapid evolution from a gluon dominated
initial state to a quark dominated plasma close to $T_c$.
The quark to gluon ratio can be modified by about a factor of $\sim 20$ in the bulk of the
system and appears to be large also in the high $p_T$ region.

We discuss how this aspect, often overflown, can be important for a quantitative study of several key
issues in the QGP physics: shear viscosity, jet quenching, quarkonia suppression. 
Furthermore a bulk plasma made by more than
$80\%$ of quarks plus antiquarks provides a theoretical basis for hadronization via quark coalescence.

\end{abstract}


\maketitle

The search for the Quark-Gluon Plasma (QGP) started its golden age thanks to the experiments
at the Relativistic Heavy Ion Collider (RHIC) that have supplied
convincing physical evidences that a new state of matter has been created \cite{Adams:2005dq,Adcox:2004mh}. Such a matter has
a very small shear viscosity \cite{Romatschke:2007mq,Song:2007ux,Song:2010mg}, a high opacity to high$-p_T$ particles \cite{Gyulassy:2004zy}, 
a strong screening of the interaction able to significantly dissociate charmonia \cite{Rapp:2008tf}, and exhibits a modification
of the hadronization respect to the vacuum toward a quark coalescence mechanism \cite{Greco:2003mm,Fries:2003kq,Fries:2008hs}.
Furthermore some RHIC data hints to the creation of an "exotic" initial state of matter,
the Color Glass Condensate (CGC), that could be the high-energy limiting state of the QCD interaction
\cite{Gyulassy:2004zy}.
The new and upcoming experiments at the Large Hadron Collider (LHC) have confirmed the main
gross properties observed at RHIC \cite{ALICE-QM}, but they will allow to explore a larger temperature range
also with quite different heavy-quark abundancy and will provide more suitable
conditions for creating CGC phase as initial state. 

The several probes mentioned above  rely on the comparison between experimental data and
model predictions. A closer look into the several theoretical approaches to the different
QGP probes reveals that in some cases the QGP is described as a Gluon Plasma.  Indeed this initially
should be the case because most of the particles come from low $x$ momentum fraction where the
nucleon parton distribution functions are gluon dominated. Hence, for example, is the case of the 
most popular jet quenching models assuming a bulk gluon matter.
In other cases as in the viscous hydrodynamics
a chemical quark-to-gluon equilibration is implicit
in the employment of a lattice QCD (lQCD) equation of state $P(\epsilon)$.
For the study of quarkonia instead usually one considers a plasma of gluons or an equilibrated QGP 
according to a massless quark-gluon description acting for dissociation.
On the contrary the observation of quark-number scaling in the elliptic flow and the
baryon over meson enhancement are explained by quark coalescence models 
based on a quark dominance in the plasma \cite{Greco:2003mm,Fries:2008hs}.

Certainly despite a lack of full integration of the different
descriptions of the QGP, all of these have been useful simplifications that allowed to successfully
identify the creation of the QGP plasma and its gross properties.
Nonetheless once we have identified the main qualitative features of the matter created
in uRHIC's a quantitative knowledge of properties like the $\eta/s$ or
the solution to open issues on the jet quenching mechanism (radiative vs collisional), quarkonia dissociation-regeneration, 
hadronization mechanism and existence of a CGC matter, requires to consider the poorly explored
issue of the "chemical" composition of the QGP. 

The assumption of chemical equilibrium of the QGP, when considered, is usually discussed 
treating the QGP as a gas of massless quarks and gluons; therefore the expected ratio is given
simply by the ratio of the degrees of freedom $N_{q+\overline{q}}/N_g=d_{q+\overline{q}}/d_g=
2N_c\,N_f/(N_c^2-1)=9/4$ for a system with 3 flavors.
On the other hand, as well known from lQCD, the QGP appears to be significantly different 
from a mere massless gas, showing deviation of both the energy density and the 
pressure from the $\epsilon/T^4=3P/T^4 =cost.$, and in particular exhibiting a large trace anomaly $\langle\Theta_\mu^\mu\rangle=\epsilon - 3 P$.
It has been shown that such a behavior can be described in terms of a massive quasiparticle model
in which both gluons and quarks acquire a thermal mass $m(T) \sim g(T)\, T$, as suggested also
by the Hard Thermal Loop (HTL) 
approach \cite{HTL1,HTL3,HTL4,HTL7} or
dimensionally reduced screened perturbation theory (DRSPT) \cite{DR1,DR2} or HTLpt \cite{Andersen_HTLpt}.
This comes out also  from extracting a gluon propagator from lattice results in the Landau gauge \cite{Oliveira:2010xc},
or using the pinch technique \cite{Binosi:2009qm} or a T-matrix approach from lQCD free-energy \cite{Lacroix:2012pt} .
All such approaches suggest a finite $m_g \sim 0.5 - 1 \rm GeV$, 
a value quite close to the one obtained by fitting a QPM to lQCD thermodynamics, as done in the present work.

A quasi-particle model (QPM) with a $T-$dependent Bag constant has been successfully applied to quantitatively describe lQCD results for equilibrium thermodynamics \cite{Plumari:2011mk,qpm,Ratti:2011au}
including the recent ones performed with an unprecedented level of accuracy
at the physical quark masses \cite{Borsanyi:2010cj}. 
It is also interesting to note that QPM is able to correctly 
predict $\eta/s\leq$ 0.2 close to $T_c$ \cite{Plumari:2011mk,Cassing} with quasi-particle widths still significantly smaller than
the mass itself \cite{Cassing}. We neglect in the present work the finite width which can be expected to marginally affect the quark
to gluon ration respect to Eq.(\ref{eq-ratio}).

Moreover lQCD calculations on charge-charge correlations show that up,down and strange charges are transported
as single charges and off-diagonal elements disappear above $\rm T_c$ indicating that quark and gluon
quasi-particles are still good quantum numbers at least not too close to $\rm T_c$.

We notice that
if the QGP can be described in terms of finite mass excitations this has a strong impact on the
quasiparticle chemical ratio $N_{q+\overline{q}}/N_g$. 
In fact at equilibrium one has:
\begin{eqnarray}
\frac{N_{q+\overline{q}}}{N_g}=\frac{d_{q+\overline{q}}}{d_g}
\frac{m_q^2(T)\, K_2(m_q/T)}{m_g^2(T)\, K_2(m_g/T)}\, ,
\label{eq-ratio}
\end{eqnarray}
where $K_2$ is the Bessel function and $m_{q,g}(T)$ are the $T-$dependent quark and gluon masses that can be determined by a fit 
\cite{Plumari:2011mk} to recent lQCD calculations \cite{Borsanyi:2010cj}.
In Fig.\ref{eq-ratio}, we show by solid line the equilibrium ratio when the fit to lQCD 
$\epsilon(T)$ is done assuming $m^2_q/m^2_g=3/2 \cdot (N_c^2-1)/N_c(2N_c+N_f)=4/9$
according to a pQCD scheme \cite{qpm}. 
We plot also by dash-dotted line the expected ratio
assuming the HTL  ratio $m^2_q/m^2_g=1/9$ ratio. This of course leads to a larger
quark abundancy due to larger difference between quark and gluon masses. 
Furthermore in Ref. \cite{Plumari:2011mk} some of the authors have shown that in 
the last case one can better describe the diagonal quark susceptibilities, in the following
we will consider the more commonly assumed pQCD case.
This  would also prevent from overestimating the magnitude of the effect discussed.

\begin{figure}[htbp]
\begin{center}
\vspace{0.55cm}
\includegraphics[scale=0.3]{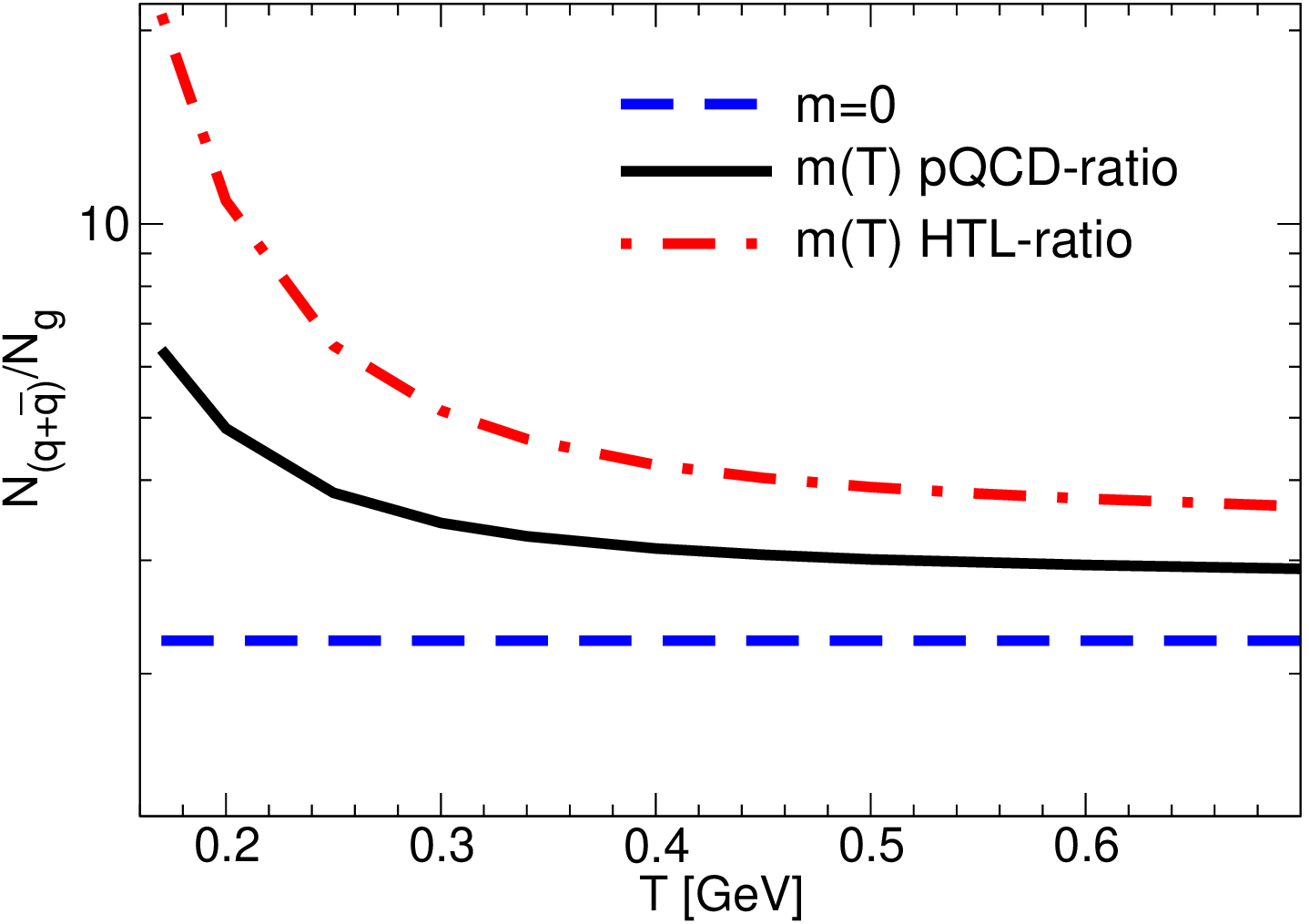}
\caption{\label{nqng-qpm}
Quark to gluon ratio at equilibrium
as a function of the temperature as predicted by QPM \cite{Plumari:2011mk}. 
Solid line is assuming a $m_q/m_g$
ratio according to pQCD and dot dashed line according to HTL; by 
dashed line it is indicated the massless case.}
\end{center}
\end{figure}

In this Letter we discuss the issue of the quark-to-gluon ratio of the matter created in uRHIC's
at both RHIC and LHC energies.
We employ a Boltzmann-Vlasov transport theory to simulate the partonic stage of the HIC in a realistic way.
In the last years several codes have been developed based on transport theory at the cascade level
\cite{Zhang:1999bd,Molnar:2001ux,Xu:2004mz,Ferini:2008he},
i.e. including only collisions between massless partons, with quite rare exceptions
\cite{Cassing:2009vt,Plumari:2010}. These approaches have been more recently developed to
study the dynamics of the partonic stage of the HIC at fixed shear viscosity 
\cite{Ferini:2008he,Greco:2008fs,Plumari:2010,Plumari:2011re} with the advantages
to explore possible effects of kinetic non-equilibrium having also
a wider range of validity in $p_T$ and in $\eta/s$. 
 
We present here for the first time results within a transport approach that includes
the mean field dynamics associated to the thermal self-energies generating the finite
mass $m(T)$ in the QPM discussed above \cite{Plumari:2011mk,qpm}. 
The approach is formally similar to the one developed in 
Ref. \cite{Plumari:2010} for the NJL mean field dynamics, but here the quasiparticle 
mean field allows to account for the proper equation of state, $P(\epsilon)$,
as evaluated from lQCD.
In such a picture the relativistic Boltzmann-Vlasov equation can be written as follows:
\begin{equation}
 [p^\mu \partial_\mu+ m^*(x) \partial_\mu m^*(x)\partial ^{\mu}_p]f(x,p)=
{\cal C}[f](x,p)
\label{BV-equation}
\end{equation}

where $\mathcal{C}(x,p)$ is the Boltzmann-like collision integral, main ingredient of the several cascade
codes:
\begin{equation}
{\cal C} \!=\! 
\int\limits_2\!\!\! \int\limits_{1^\prime}\!\!\! \int\limits_{2^\prime}\!\!
 (f_{1^\prime} f_{2^\prime}  -f_1 f_2) \vert{\cal M}_{1^\prime 2^\prime \rightarrow 12} \vert^2 
 \delta^4 (p_1+p_2-p_1^\prime-p_2^\prime)
\label{coll-integr}
\end{equation}
where $\int_j= \int_j d^3p_j/(2\pi)^3\, 2E_j $, $f_j$ are the particle distribution functions,
while ${\cal M}_{f\rightarrow i}$ denotes the invariant transition matrix for elastic 
as well as inelastic processes.  
The elastic processes have been implemented and discussed in several previous
works \cite{Zhang:1999bd,Molnar:2001ux,Xu:2004mz,Ferini:2008he}.
The inelastic processes between quarks and gluons ($gg \leftrightarrow q\overline{q}$)
is instead the main focus of the present Letter.
We have evaluated the matrix element in a pQCD LO order scheme. The tree diagrams contributing to the 
$gg \leftrightarrow q\overline{q}$ correspond to the $u,t,s-$channels: ${\cal{M}}={\cal{M}}_s+{\cal{M}}_t+{\cal{M}}_u$. 
For the massless case the cross sections for such
processes are the textbook pQCD cross section for jet production in high-energy proton-proton 
collisions. With massive quarks the calculations are the renowned Combridge cross
sections \cite{Combridge:1978kx} used to evaluate heavy quark production.
In our case we have considered a finite mass for both gluons and quarks together with a dressed
gluon propagator.
The details of the calculations are lengthy and will be published elsewhere \cite{scardina},
but they are quite similar to the one in \cite{Biro:1990vj} for finite current strange quark mass.
We only mention that the cross section is dominated by the 
$t-$ and $u-$channel and their interference while the $s-$channel alone is
negligible. The squared matrix element of the $t-$channel is given by:
\begin{eqnarray}
|(t-m^{2}_{q})\,{\cal{M}}_t|^2 &=& \frac{8}{3} \pi^2 \alpha^2_s [(m^2_q-t)(m^2_q-u) \nonumber \\
& & -2m_q^2(t+m^2_q)- 4\,m^2_q m_g^2 - m_g^4]
\label{matrix}
\end{eqnarray}
and of course by crossing symmetry the $u-$channel, can be obtained by $u\leftrightarrow t$
exchange. 

We have employed a running coupling $\alpha_s(Q)$ according to first order
expansion and the energy scale given by $Q^2=(\pi \, T)^2$.
The pQCD scheme with
renormalized fermionic and bosonic lines should provide a useful
guideline for the estimate of the pertinent cross section, keeping in mind that it can be expected
that the real cross section could be even larger than the estimated ones, as for
the elastic scattering processes. 
However we will see that the evaluated cross sections are already large enough that variations within
a factor of two only marginally affects the final quark to gluon ratio,
in fact the system gets anyway quite close to equilibrium.

The thermodynamical self-consistency of the QPM requires a self-consistency
between the Bag constant and the effective mass of the quasiparticles \cite{Plumari:2011mk}
which leads to a gap-like equation coupled to Eq.(\ref{BV-equation}):
\begin{equation}
 \frac{\partial B}{\partial m_i} =-
 \int \frac{d^{\,3} \vec{p}}{(2\pi)^3 } \frac{m_i(x)}{E_i(x)}  f_i(x,p) \, 
\label{gap-equation} 
\end{equation}
with $i=q,\overline q,g$.
Eq.(\ref{gap-equation}) allows to evaluate locally the mass in Eq.(\ref{BV-equation}) also in non-equilibrium conditions guaranteeing the conservation of the energy-momentum tensor of the fluid.
For the numerical solutions of Eqs.(\ref{BV-equation}) and (\ref{gap-equation}) we use a three dimensional 
lattice that discretizes the space and the standard test particle methods that samples the distribution function 
$f(x,p)$ by means of an ensemble of points in the phase-space, for more details see Ref. \cite{Plumari:2010}.
\begin{figure}[htbp]
\begin{center}
\vspace{0.6cm}
\includegraphics[scale=0.3]{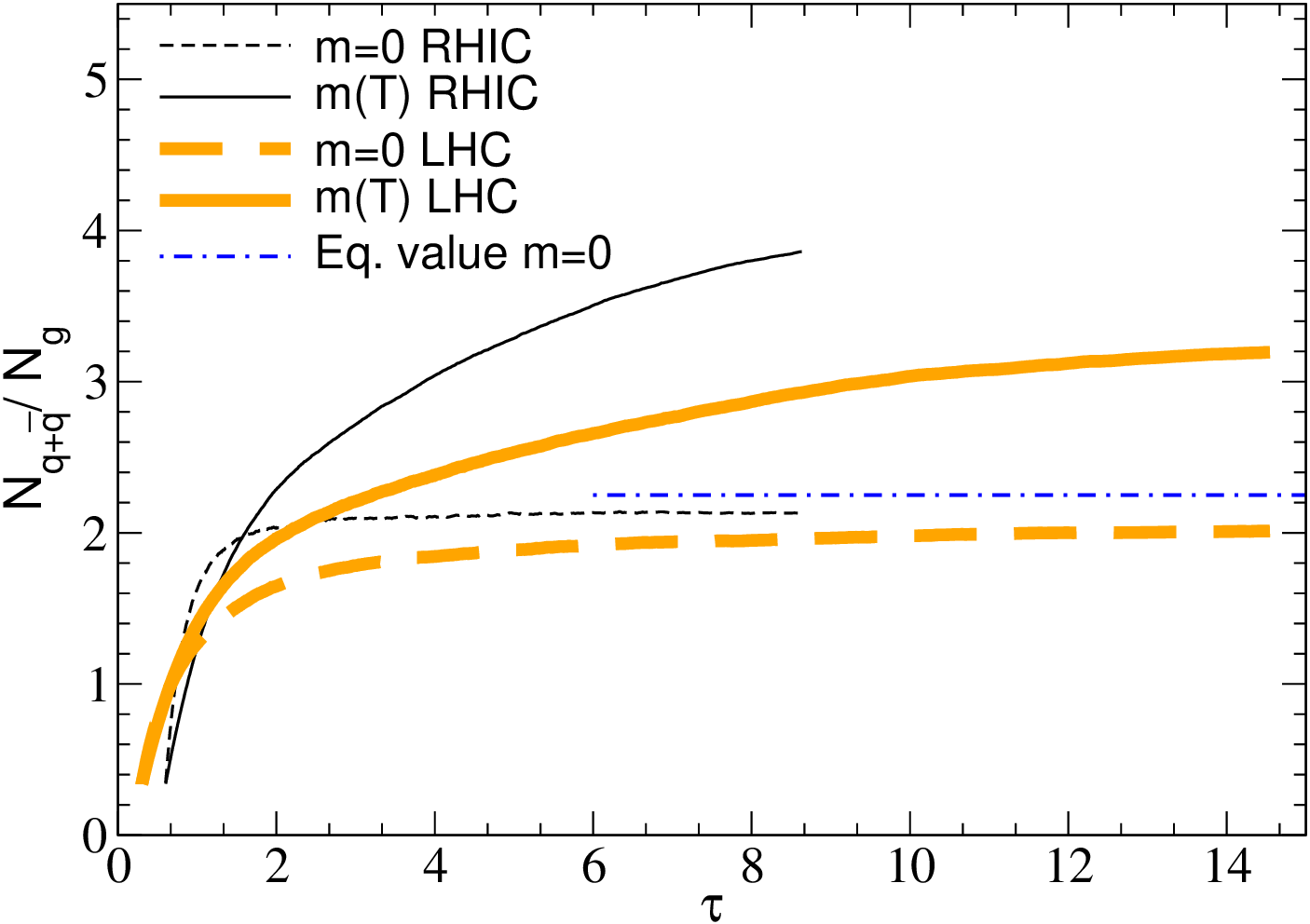}
\caption{\label{ratio-time} 
Quark to gluon ratio as a function of time  in
$Au+Au$ at $\sqrt{s}= 200 \rm AGeV$ (black thin lines) and for $Pb+Pb$ 
at $\sqrt{s}= 5.5 \rm ATeV$ (light thick lines).
Dashed lines are for the massless case and the solid for the massive case}
\end{center}
\end{figure}
We have carefully checked that the numerical implementation 
in a box is able to reproduce correctly both the kinetic 
equilibrium and in particular the chemical one, i.e. 
the right abundancy of quarks, antiquarks and gluons according to 
Eq.(\ref{eq-ratio}), which implies also a proper implementation of the detailed
balance.

For the numerical implementation of Eqs.(\ref{BV-equation}) and (\ref{gap-equation})
we discretize the coordinate space using a three dimensional lattice, as described in Ref.
\cite{Ferini:2008he,Xu:2004mz}. In particular, using the standard test particle method the distribution 
function $f_j$ is sampled in the phase-space by mean of an ensemble of $A_j=N_{test} \cdot N_{real}$ points, 
for each flavor $j$, with $N_{real}$ the real particle number associated to $f_j$. 
In such a way the solution of the transport equations is equivalent to solve the 
Hamilton equations for the test particles, which can be written 
in the following form:
\begin{eqnarray}
\label{Hamiltons}
\boldsymbol{p}_i(t^+)&=&\boldsymbol{p}_i(t^-)-2 \, \delta t \, \frac{m_{j}(\boldsymbol{r}_i,t)}{E_i(t)} \vec{\nabla}_r m_{j}(\boldsymbol{r}_i,t) + coll. \nonumber \\
\boldsymbol{r}_i(t^+)&=&\boldsymbol{r}_i(t^-)+2 \, \delta t \,\frac{\boldsymbol{p}_i(t)}{E_i(t)}
\end{eqnarray}
with $j=q,\bar{q},g$ and $t^\pm=t \pm \delta t$, being $\delta t$ the numerical mesh time.
The term $coll.$ on the right hand side of Eq.(\ref{Hamiltons}) 
indicates the effect of the collision integral, as described by Eq.(3)
and numerically solved as in Ref.\cite{Xu:2004mz,Ferini:2008he}.
For thermodynamical consistency 
Eq.s(\ref{Hamiltons}) are coupled to the gap-like equation (\ref{gap-equation}) that
discretized on the lattice and for point-like test particles becomes:
\begin{eqnarray}
\label{NumericalGap}
\frac{\partial B}{\partial m_{j}}= - \frac{m_j}{N_{test}\, \Delta V_\alpha} \sum_{i=1}^{A_{j}} \frac{1}{E_i(m_{j})},
\end{eqnarray}
where $\Delta V_\alpha=\tau A_T  \tanh \eta_\alpha$ is the volume of 
each cell of the space lattice, being $A_T=0.25 fm^2$ the area in 
the transverse direction and $\eta_\alpha$ the space-time rapidity 
of the center of the cell. The term ${\partial B}/{\partial m_{j}}$ is 
given by the quasi-particle model  \cite{Plumari:2011mk} according to the parametrization
corresponding to the lQCD equation of state of Ref.\cite{Borsanyi:2010cj}.
The coupled eqs.(\ref{Hamiltons}) and (\ref{NumericalGap}) are solved in a self-consistent way and the procedure is reiterated 
at each time steps. We have checked that in a box the numerical solution corresponds to
a system at equilibrium with the pressure and the energy density
corresponding to the lQCD equation of state as a function of the temperature.

We have simulated $Au+Au$ collisions at $\sqrt{s}= 200$ AGeV and $Pb+Pb$ at $\sqrt{s}= 5.5 \rm ATeV$
for $0-10\%$ centrality.
The initial conditions in the $r$-space are given by the standard Glauber condition. In the $p$-space,
we use a Boltzmann-Juttner distribution function up to a transverse momentum $p_T=2$ GeV and
at larger momenta mini-jet distributions are implemented, as calculated by pQCD at NLO order \cite{Greco:2003mm}. 
At RHIC the initial maximum temperature at the centre of the fireball is $T_0=340$ MeV and
the initial time $\tau_0=0.6$ fm/c as in successful hydrodynamical simulations
(corresponding also to the $\tau_0 \cdot T_0 \sim 1$ criterium). 
For $Pb+Pb$ collisions at $\sqrt{s}= 5.5 \, \rm ATeV$, we have 
$T_0=600$ MeV and $\tau_0\sim 1/T_0=0.3$ fm/c.
In Fig.\ref{ratio-time} it is shown the time evolution of the ratio $R_{qg}=N_{q+\overline{q}}/N_g$
for both the massless case (dashed lines) and the massive quasi-particle
case (solid lines) for both RHIC and LHC energies. We clearly see that in the massless case the system reaches very quickly,
in less than 1 fm/c, the chemical equilibrium given by $R_{qg} \sim 2$. 
Therefore in the massless case to assume a chemical equilibrium in modeling
the plasma composition can be already considered a reasonable approximation.

When the quasi-particle massive case is considered, we can again see that one can reach
quickly a value of $R_{qg} \sim 2$, then there is a slower rise with a continous
increase up to $R_{qg} \sim 4$ for $Au+Au$ .
We notice that for the massive case the equilibrium value is strongly $T$ dependent, especially
close to $T_c$ (see Fig.\ref{nqng-qpm}), and the system more dilute in this stage
is not capable to follow such a rapid evolution. Nonetheless, we find that the fireball reaches a value relatively
close to the equilibrium at $T \sim T_c$ 
and eventually it is composed by about $80 \%$ of quark plus anti-quarks.
At LHC the trends are very similar but a longer part of the lifetime is spent in a $T$-region
where the equilibrium $R_{qg}$ is nearly constant. This results into
a moderately smaller final ratio.

We show in Fig.\ref{ratio-time} the result starting from an initial gluon dominated plasma
with a $R_{qg} = 0.25$, however changing
this initial ratio by a factor of two affects the final ratio by less than a $10 \%$, while
the effect we are describing modifies the value of $R_{qg}$ by more than an order of magnitude.  

\begin{figure}[htbp]
\begin{center}
\vspace{0.55cm}
\includegraphics[scale=0.3]{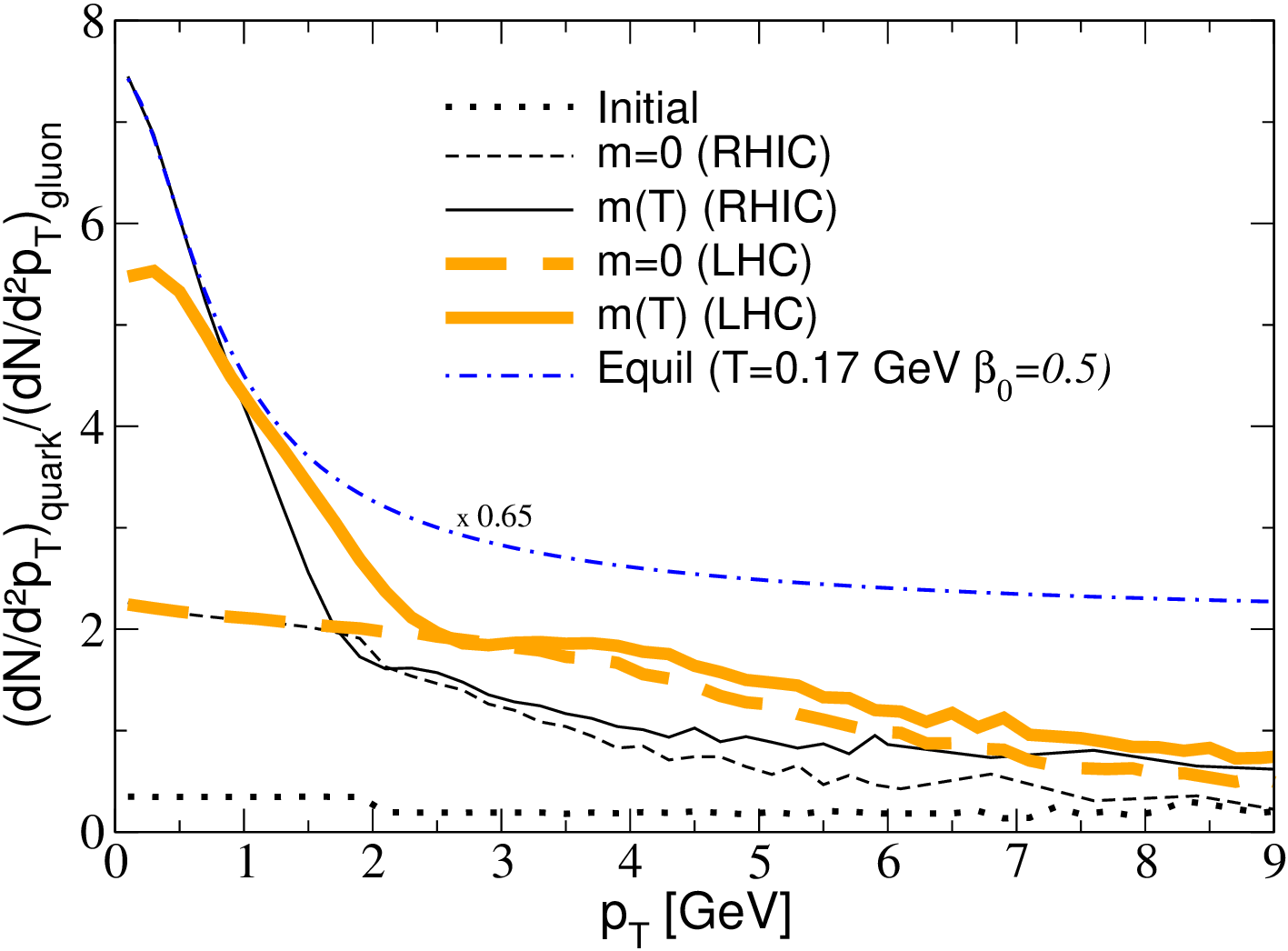}
\caption{\label{ratio-pt}
Quark to gluon ratio vs $p_T$ at the freeze-out time.
Labels as in Fig.\ref{ratio-time}. The thin dot-dashed line represents the full equilibrium
ratio, Eq. (\ref{ratio-pt-anal}); see text for details. }
\end{center}
\end{figure}

In Fig.\ref{ratio-pt} the $p_T$ dependence of the quark to gluon ratio is shown for 
the initial distribution (dotted line) and the freeze-out distributions: massless
case (dashed lines) and massive case (solid lines). Thin black lines are for $Au+Au$
and thick light ones fore $Pb+Pb$.
We see the large difference between the massless and the massive case and
also that the net gluon to quark conversion extends up to quite large $p_T$. 
The decrease of the ratio with $p_T$ can be expected considering that high$-p_T$ particles
can more easily elude the equilibration dynamics.
However, in the massive case, we note a quite strong dependence below $p_T=2$ GeV that
has not to be interpreted as a fast detachment from the chemical equilibrium.
In fact the $p_T-$dependence of the ratio can be evaluated analytically at
equilibrium and it is given by
\begin{equation}
 \frac{dN/d^2\,p_T|_{q+\overline{q}}}{dN/d^2\,p_T|_g}=\frac{d_{q+\overline{q}}}{d_g}
\frac{m_{T}^{q} e^
{-\gamma[(m_{T}^{q}-\beta_0 p_T)/T]}}{m_{T}^{g} e^
{-\gamma[(m_{T}^{g}-\beta_0 p_T)/T]} }
\label{ratio-pt-anal}
\end{equation}

where $\beta_0$ is the radial flow velocity, $m_{T}^{q}$ and $m_{T}^{g}$
are the transverse masses.
We plot in Fig.\ref{ratio-pt} by thin dot-dashed line such a function 
rescaled by a factor 0.65 accounting for the lack of full thermalization.
The strong $p_T$ dependence
obtained in the transport simulation follows very closely the equilibrium behavior
at least up to $p_T \sim 1.5$ GeV.
This is a well known effect predicted by hydrodynamics and experimentally observed 
from SPS to LHC energy for hadronic spectra. 

Our study shows that the most common microscopic description of the lQCD thermodynamics 
implies a plasma that in uRHIC's evolves from a Glasma toward a quark dominated plasma.
Such an effect appears to be quite solid despite uncertainties in the inelastic cross sections
and could be even larger and robust
if one considers that the quark susceptibilities computed on lQCD
hints at even smaller $m_q/m_g$ ratio \cite{Plumari:2011mk} and hence quite larger 
equilibrium value of $R_{qg}$, see dashed-dotted line in Fig.\ref{nqng-qpm}. 
Therefore the evolution of quark composition of the QGP should not be discarded for any quantitative
studies of the several probes of the QGP {and its properties and it is a direct implications
of the many studies on lQCD thermodynamics based on quasi-particles approaches.} 

Our result provides also a support to the quark coalescence hadronization mechanism,
capable to explain two main observations at RHIC: the baryon/meson anomaly and the quark number scaling 
of the elliptic flow \cite{Fries:2008hs}. 
One of the main criticism to the coalescence model is the assumption of a massive quark plasma, 
which may appear to be unlikely due to the preconception of a gluon dominated
plasma in uRHIC's. 

The impact however of the chemical evolution discussed in this Letter appears to be much wider.
In fact another key physical issue is the determination of the
shear viscosity. Recently it has been pointed out that a two-component system cannot be treated 
as an effective one-component system with an average $\eta/s$ from
mixture, but one has to solve hydrodynamics for each component \cite{El:2011cp}. 
Therefore the relative abundancy of the two components can significantly affect the
evaluation of the $\eta/s$.
Another important issue is related with the particle production at high $p_T$. Our work shows
that a significant gluon to quark conversion happens also at $p_T \geq5$ GeV in the region 
where the dominant hadronization mechanism should be independent fragmentation.
In this region, even if far from chemical equilibrium, we still find a modification of $R_{qg}$
of more than a factor of 3 respect to modelings discarding the $q\leftrightarrow g$ conversion mechanism.
\begin{figure}[htbp]
\begin{center}
\vspace{0.6cm}
\includegraphics[scale=0.3]{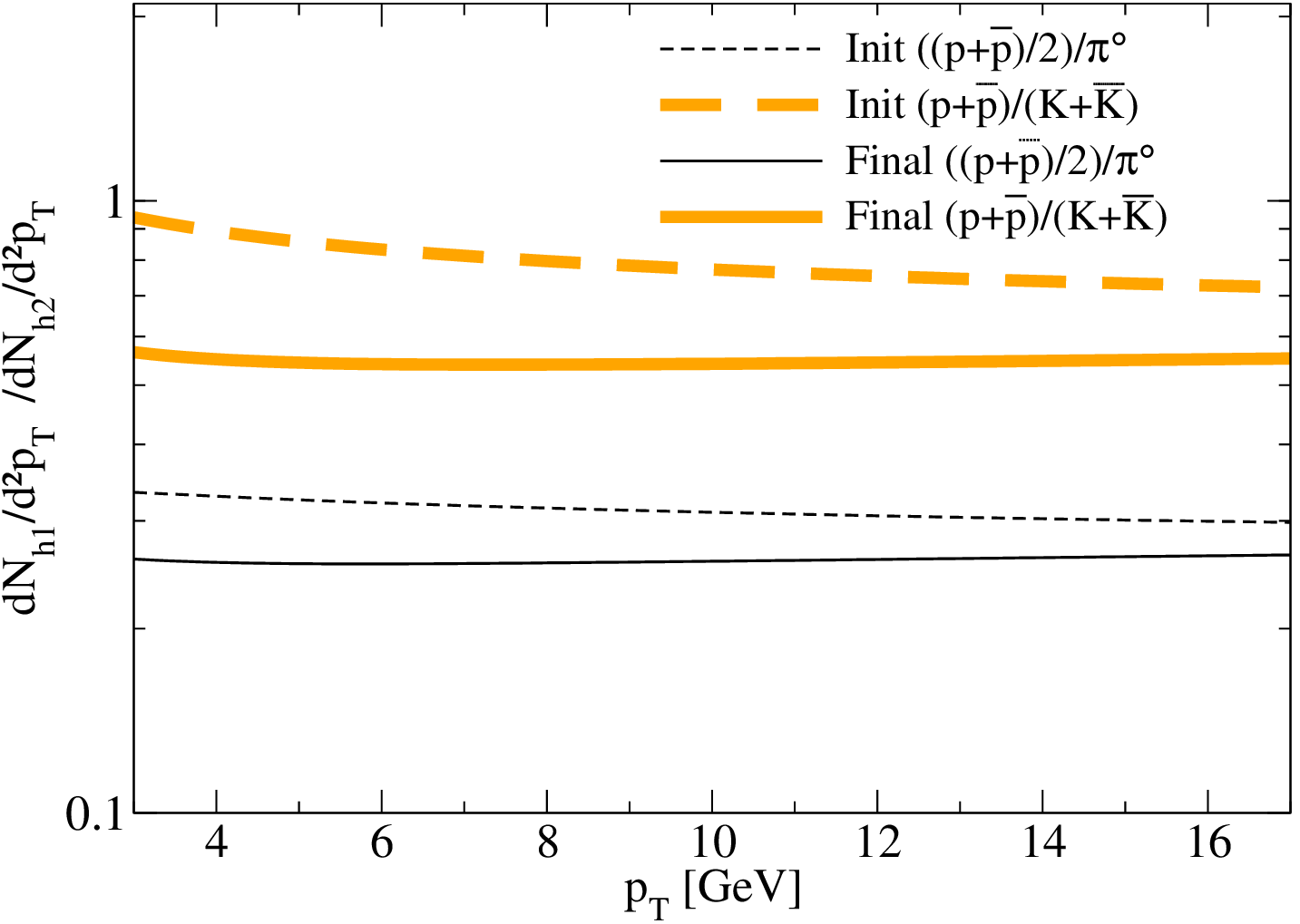}
\caption{Ratio of identified particles from independent fragmentation by AKK parametrizations.
The dashed lines are the ratio according to the initial quark and gluon distributions,
the solid lines are after the "chemical" evolution. }
\label{ratio-frag}
\end{center}
\end{figure}
In Fig.\ref{ratio-frag} , as example, we show how the ratios, $(p+\overline{p})/2\pi^0$, $(p+\overline{p})/(K+\overline{K})$, 
of identified particle can be modified when
the gluon conversion mechanism is included. 
This kind of observable asks also for a better knowledge 
of the fragmentation function (FF) in the relevant region of parton momentum fraction $x$ and $Q^2$.
Here we have used the AKK FF \cite{AKK08} just to provide a quantitative example of the potential impact of our study. 

For high$-p_T$ physics a further
important consequence could be the impact on the evaluation of the elastic scattering
energy loss and the gluon radiation mechanism responsible for the jet quenching.
In fact our study leads to a strong decrease of the relevance of $gg$ scatterings and an increase
of the $qq$ and $q\overline{q}$ ones that however are known to have quite smaller cross sections
by a factor $4-8$ depending on the specific channel.
Similarly the chemical composition of the QGP affects the physics of
quarkonia dissociation in medium. 
Indeed a preliminary study of this aspect 
has been very recently performed in Ref.\cite{Brezinski:2011ju} for bottomonium.

In conclusions our study shows that one could expect that the QGP created 
in uRHIC's, even if it is initially a Glasma should very quickly evolve 
into a plasma dominated by quark plus antiquarks close
to the cross-over temperature $T_c$. 
The results are quite robust and developments of
the QPM \cite{Plumari:2011mk,qpm} or inclusion of three-body inelastic scatterings 
may even make the effect larger and faster. 
Certainly different approaches \cite{HTL7,Andersen_HTLpt,DR1, Oliveira:2010xc,Binosi:2009qm,Lacroix:2012pt}
would have differences in the detailed behavior of $m_{q,g}(T)$, however, being $m_q/m_g Å\sim 1.5-2$,
our study suggests a rapid change in HIC from a glasma matter into a massive quark plasma close to
the equilibrium conditions implied by the several quasi-particle approaches employed for mimic the QCD thermodynamics
and studying the nuclear matter phase diagram.
The result also supplies a theoretical justification of the \textit{massive-quark} coalescence hadronization
models able to solve several puzzling observations at RHIC and LHC; moreover it 
can have a wide impact on the main physical issues of the QGP physics as a quantitative estimate
of the $\eta/s$, the study of the identified high-$p_T$ particles and the related jet quenching mechanism
as well as on the physics of the quarkonia suppression.
In general a quark dominance along with a small $\eta/s$ and a large opacity to high-$p_T$
mini-jets shifts the interpretation of the QGP toward even stronger deviations form a perturbative
behavior \cite{Shuryak:2008eq} and increased relevance of gluon-radiative energy loss \cite{Wicks:2005gt}.

\textit{Acknowledgments-}
We acknowledge the ERC support under the QGPDyn Starting Grant.

\end{document}